\documentclass[twocolumn,showpacs,preprintnumbers,amsmath,amssymb]{revtex4}
\usepackage{graphicx}
\usepackage{dcolumn}
\usepackage{bm}

\begin{document}
\title{Charge separation effects in magnetized 
       electron-ion plasma
       expansion into a vacuum}

\author{Kazumi Nishimura, Edison Liang$^{a)}$, and S. Peter Gary}
\address{Los Alamos National Laboratory,
Los Alamos, NM 87545 \\
${}^{a)}$Rice University,
Houston, TX 77005-1892}

\date{\today}
\begin{abstract}
Charge separation effects in the expansion of
magnetized relativistic electron-ion plasmas into a vacuum 
are examined using 2-1/2-dimensional
particle-in-cell plasma simulations.
The electrostatic field at the plasma surface
decelerates
electrons and accelerates ions.
A fraction of the surface electrons are trapped
and accelerated by the pondermotive force of
the propagating electromagnetic pulse,
a mechanism we call
the DRPA (diamagnetic relativistic pulse accelerator).
This charge separation is enhanced as the initial plasma
temperature is decreased.
The overall energy gain of the plasma particles 
through the
expansion strongly depends on the initial plasma temperature.
Moreover, the
electrons become relatively less energized
and the ions more energized
as the plasma temperature decreases.
\end{abstract}

\maketitle

Several kinds of plasma expansions have been investigated
over the past few decades. A model of free plasma expansion
into a vacuum is interesting for studies related to
experiments of ion jets\cite{dena79,mora03}.
Structures and instabilities of
plasmas expanding into a vacuum 
which contains a uniform ambient magnetic
field have also been examined by theoretical and
numerical methods.
Those studies are relevant
to laser-plasma
experiments\cite{okad81,ripi87}
and space plasma phenomena\cite{sydo83,wins88,sgro89}.
Also plasma injection into a magnetic field has been
theoretically investigated\cite{pete82}.

Recently, Liang {\it et al.}\cite{lian03,lian032}
have studied a new type of plasma expansion,
in which a magnetized relativistic plasma freely
expands into a vacuum with no external magnetic
field.
This model of expansions potentially can be applied
to a variety of astrophysical phenomena, 
such as gamma-ray bursts\cite{lian032} or
astrophysical jets.
When relativistic plasmas
which contain a strong transverse 
magnetic field are suddenly
released, an EM pulse forms and
begins to
propagate into the vacuum.
Simultaneously, a portion of the surface plasma 
is trapped and accelerated
by the EM pulse via the pondermotive force.
Through the expansion, the field energy of the EM pulse
is converted into the directed kinetic energy
of the surface particles.
This new mechanism of plasma acceleration is called the
diamagnetic relativistic pulse accelerator (DRPA)
\cite{lian03}.

Electrons and ions behave differently in the expansion process.
Because of their much smaller mass, the surface electrons
soon outrun the ions
and charge separation occurs.
In this Letter, 
using the results of plasma full particle simulations,
we discuss charge separation effects on
plasma energization in the relativistic
magnetized plasma expansion into a vacuum.
%

In our particle simulation code\cite{nish03},
we use a two-and-a-half-dimensional
explicit simulation scheme
based on the particle-in-cell method
for time advancing
of plasma particles.
The grid
separations are uniform, $\Delta x=\Delta z=\lambda _{e}$,
where $\lambda _{e}$ is the electron inertial length defined
by $c/\omega_{pe}$
($c$ is the speed of light and the electron plasma frequency
is $\omega_{pe}=\sqrt{e^{2}n_{0}/\epsilon _{0}m_{e}}$;
$e$ is elementary charge, $n_{0}$ is the initial electron
density, and $\epsilon _{0}$ is the dielectric constant of
vacuum).
The ratio of the electron Larmor frequency to the electron
plasma frequency 
$\Omega_{e}/\omega_{pe}$
(where $\Omega_{e}=eB_{0}/m_{e}$; $B_{0}$ is the
background magnetic field)
is set equal to 10 in these simulations.
The ion-electron
mass ratio is $m_{i}/m_{e}=100$ and the charge ratio
is $q_{i}/e=1$.
The simulation domain on the $x-z$ plane is
$-L_{x}/2\leq x\leq L_{x}/2$ and
$-L_{z}/2\leq z\leq L_{z}/2$.
To prevent large violations of Gauss's law 
due to numerical noise,
Marder's method for the electric field correction\cite{mard87} 
is adopted at every time step in the code.

We use a doubly periodic system in $x$ and $z$ directions,
and the system length is $L_{x}=240\Delta x=2400c/\Omega_{e}$
and $L_{z}=120\Delta z=1200c/\Omega_{e}$.
Initially, the electron (ion)
distributions are assumed
to be a relativistic (non-relativistic) Maxwellian respectively
with spatially uniform temperature.
The spatial distribution of the initial plasma has a
slab form with the lengths $6\Delta x\times 120\Delta z$, 
and the
plasma slab is located in the center of the system. The
background magnetic field $\mbox{\boldmath$B$}_{0}=(0,B_{0},
0)$ initially exists only inside the plasma.
Since the system length in the $x$ direction is enough large,
no plasma particles and waves can cross the $x$ boundaries
during this calculations (until $t\Omega_{e}=1000$).

We describe here the results of two simulations; one with
initial temperatures of $k_{B}T_{e}=k_{B}T_{i}=5 MeV$ 
and one with initial
temperatures of $k_{B}T_{e}=k_{B}T_{i}=100 keV$. 
The qualitative
evolution of the $5 MeV$ temperature case is
illustrated in Figure 5 of Ref.\cite{lian03}.
A pulse of fast electrons moves
rapidly away from the ions in the $x$-direction, 
carrying an imbedded
electromagnetic pulse ($B_y > 0$, $E_z < 0$) with them which permits
them to $E\times B$ drift.  Behind this pulse
the $B_y$ and $E_z$ are relatively weak, and the expanding ions lag
well behind the electron pulse.  This charge separation leads to
an $E_x$ in the region between 
the electron pulse and the fastest ions.

\begin{figure}
\includegraphics[width=8cm]{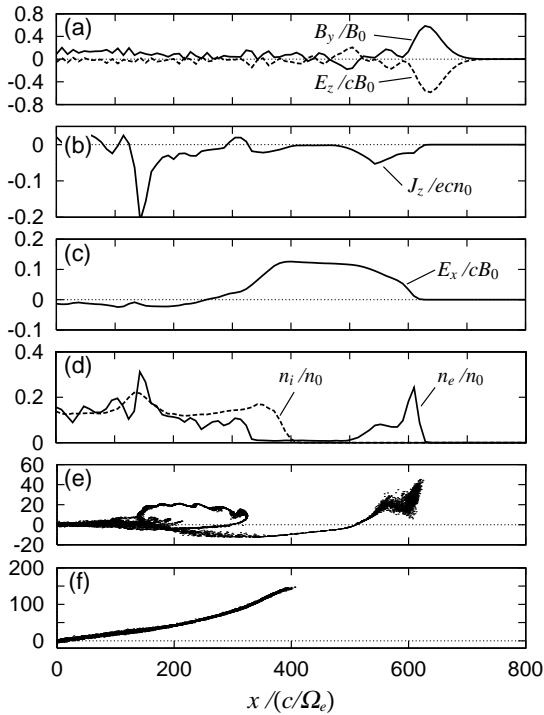}
\caption{\label{flds-x}
         Simulation results for $k_{B}T=100keV$ plasma
         at $t\Omega_{e}=600$.
         (a) Electromagnetic field ($E_{z}$ and $B_{y}$),
         (b) current density ($J_{z}$),
         (c) electrostatic field ($E_{x}$), and (d)
         electron and ion density ($n_{e}$ and $n_{i}$)
         as functions of $x$.
         All of the quantities are obtained at $z=0$.
         Phase plots $p_{x}/m_{e}c-x$
         for (e) electrons and (f) ions.
         These profiles should be compared with those
         of the $k_{B}T=5MeV$ case in Ref.[8].}
\end{figure}

Spatial profiles of the EM field
and the plasma density at $t\Omega_{e}=600$
are displayed in Figure~\ref{flds-x}
for the $100keV$ temperature case.
In Fig.~\ref{flds-x}(a), the EM pulse is propagating
in the positive $x$ direction.
The electrostatic field $E_{x}$ 
shown in Fig.~\ref{flds-x}(c)
is generated by the charge separation between 
electrons and ions.
As shown in Fig.~\ref{flds-x}(d), electrons are distributed
spatially more extensively
than ions because of the larger mobility of
electrons.
In the case of electron-positron expansion~\cite{lian03},
there is no difference in the density distributions of
electrons and positrons.
Consequently, both electrons and positrons are
equally energized by the DRPA.
But in the present electron-ion case,
only electrons can follow
and get energized by the EM pulse, while the ions
are accelerated in a secondary manner by the
charge-separation electric field.

Phase plots $p_{x}/m_{e}c-x$
of (e) electrons and (f) ions 
at $t\Omega_{e}=600$ in the $100keV$ temperature case
are also shown in
Fig.~\ref{flds-x}.
In Fig.~\ref{flds-x}(e), electrons are decelerated
in the region $300<x/(c/\Omega_{e})<600$,
where the electric field $E_{x}$ exists.
The surface electrons in the vicinity of
$x\simeq 600c/\Omega_{e}$ are strongly energized
by the EM pulse.
Since some of the expanding electrons are reflected
at the potential wall of the electrostatic field
at $x\simeq 300c/\Omega_{e}$,
a loop structure in phase space can be seen.
On the contrary, the front
ions in this region are accelerated by $E_{x}$ and
the averaged kinetic energy of the front ions tends to
increase with time.
The front ions are not affected by the DRPA
since the ions cannot follow the EM pulse.
\begin{figure}
\includegraphics[width=8cm]{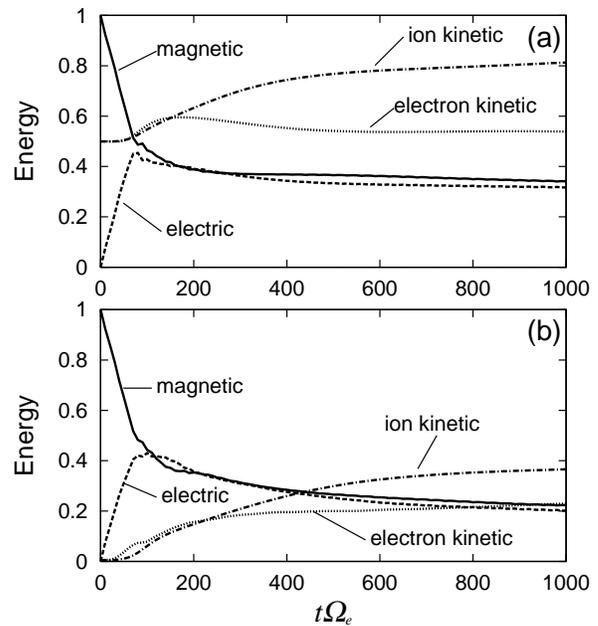}
\caption{\label{energy-t}
         System-integrated energy in the magnetic field,
         electric field, electrons, and ions for (a)
         $k_{B}T=5MeV$ and (b) $k_{B}T=100keV$ plasmas.
         Each energy is normalized by the initial magnetic
         field energy in each figure.}
\end{figure}
Hence, we obtain an interesting result regarding
particle energization in the
electron-ion plasma expansion:
the surface electrons expanding in phase with
the EM pulse are accelerated by the DRPA, but
a portion of the front
electrons are also decelerated by the electrostatic field
$E_{x}$
caused by the charge separation,
while the front ions are always accelerated by 
$E_{x}$.
Here,
we showed only the $100keV$ temperature case in Fig.~\ref{flds-x}.
See Ref.\cite{lian03} for
the profiles of the $5MeV$ case.

Temporal evolutions of each energy component
through the expansion are shown in Figure~\ref{energy-t}
for two cases with 
(a)$k_{B}T_{e}=k_{B}T_{i}=5MeV$ and
(b)$k_{B}T_{e}=k_{B}T_{i}=100keV$.
Note that in case (a) the initial plasma $\beta$
(=plasma pressure/magnetic pressure)$=0.216$
while in case (b) $\beta=0.02$, so the plasma is
strongly magnetic-dominated.
During $0<t\Omega_{e}<80$, 
roughly corresponding to the light crossing time of
the initial plasma slab,
the system evolves
from the initial magnetostatic configuration
into two counter-propagating EM pulses loaded with plasma.
In this early phase,
the magnetic energy decreases and the electric energy
increases until the two are almost equal.
After $t\Omega_{e}\simeq 80$, the field energy
decays and
is converted into the directed kinetic energy of plasma
particles.
In Figs.~\ref{energy-t}(a) and (b),
we observe the monotonic increase of the total kinetic energy
of the ions at all times.
The total electron energy, however, 
decreases from $t\Omega_{e}=80$ to 400 in the $k_{B}T=5MeV$
case, and
hardly changes
after $t\Omega_{e}=400$ in Fig.~\ref{energy-t}(a).
Even for the $100keV$ case of
Fig.~\ref{energy-t}(b), 
\begin{figure}
\includegraphics[width=8cm]{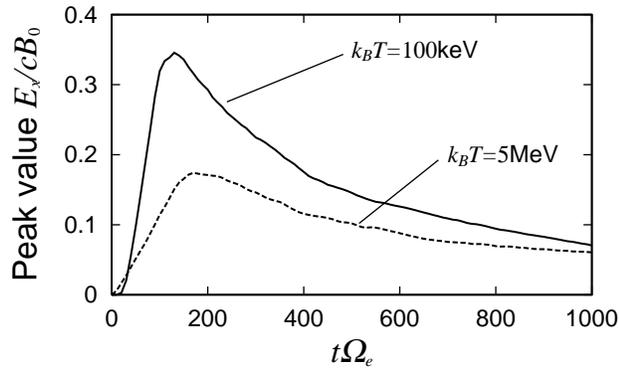}
\caption{\label{ex-t}
         Temporal evolution of peak value of electrostatic
         field $E_{x}$.
         Comparison between different temperature plasmas.}
\end{figure}
though the total electron energy increases monotonically,
we see that the rate of increase of
electron energy is much smaller than that of the ion energy.
In addition, the decrease of the total electron kinetic
energy in Fig.~\ref{energy-t}(a) is mainly caused by
the deceleration due to the electrostatic field $E_{x}$
(we described the details in Figure~\ref{flds-x}).
The same thing happens also in Fig.~\ref{energy-t}(b).
But because the initial thermal energy of electrons
is less than the directed kinetic energy given by
the DRPA, the overall kinetic energy of electrons
increases monotonically for the $100keV$ plasma.

In Figure~\ref{ex-t}, the peak value of $E_{x}$ for
different initial temperatures are shown as a function of
time.
The electrostatic field $E_{x}$ peaks
in the vicinity of the ion front.
At the early phase of the plasma expansion
($t\Omega_{e}\simeq 180$ in Fig.~\ref{ex-t}), 
the field $E_{x}$
is formed by the charge separation.
After that period, 
the density difference between electrons
and ions is gradually reduced 
by ion acceleration and electron
deceleration due to $E_{x}$.
As a result of this,
the field $E_{x}$
becomes weaker.
A plasma tends to be spatially broader as its
temperature becomes higher.
Accordingly, the higher the plasma temperature, the
smaller the density difference between electron and ion becomes.
That is why the peak value of $E_{x}$ in the case of
the $100keV$ temperature plasma tends to be 
larger than that of
the $5MeV$ temperature plasma in Fig.~\ref{ex-t}.

\begin{figure}
\includegraphics[width=8cm]{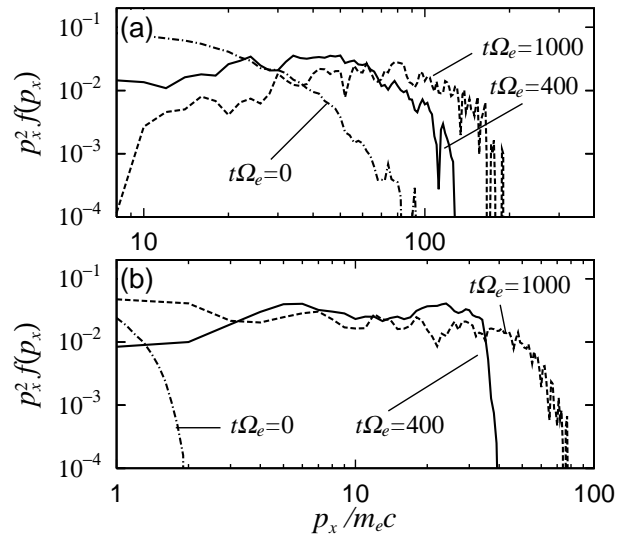}
\caption{\label{momentum}
         Momentum distributions of electrons
         in the expansion front $p_{x}/m_{e}c-x$
         for (a) $k_{B}T=5MeV$ and (b) $k_{B}T=100keV$
         plasmas at $t\Omega_{e}=0$, 400, and 1000.}
\end{figure}

Let us evaluate quantitatively
how the field $E_{x}$ influences particle energization.
Fig.~\ref{momentum} displays
the momentum distribution of the front electrons
at three different times for initial plasma temperatures of
(a) $k_{B}T=5MeV$ and
(b) $k_{B}T=100keV$.
The front electrons are accelerated by the DRPA and
simultaneously decelerated by the electrostatic field
as the expansion proceeds.
The energy gain of these electrons seems to be positive
and the kinetic energy continues to increase as time
elapses, as shown in Fig.~\ref{momentum}.
\begin{figure}
\includegraphics[width=8cm]{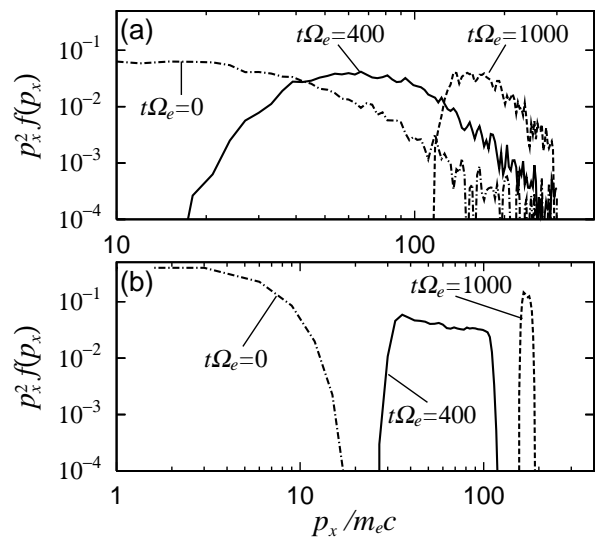}
\caption{\label{momentum2}
         Momentum distributions of ions
         in the expansion front $p_{x}/m_{e}c-x$
         for (a) $k_{B}T=5MeV$ and (b) $k_{B}T=100keV$
         plasmas at $t\Omega_{e}=0$, 400, and 1000.}
\end{figure}

This means that the acceleration process dominates 
the deceleration process overall for the front electrons.
Figs.~\ref{momentum}(a) and~\ref{momentum}(b) 
show that the electron energy gain
depends on the initial plasma temperature.
The electric field induced by
charge separation tends to be smaller when the plasma
temperature is higher.
Thus, the front electrons are more strongly decelerated
in the low temperature case, and
the energy gain of the electrons becomes large
when the temperature is high.

Figure~\ref{momentum2} displays the momentum distribution
of the ions in the expansion front
for (a) $k_{B}T=5MeV$ and (b) $k_{B}T=100keV$ 
plasmas.
Contrary to the case of electrons, the peak value of
the distribution becomes smaller as the initial temperature
increases
[The peak values are $p_{x}\simeq 140m_{e}c$ in 
Fig.~\ref{momentum2}(a) and $p_{x}\simeq 170m_{e}c$
in Fig.~\ref{momentum2}(b), respectively].
The front ions are accelerated only by the electrostatic
field of the charge separation.
The energy gain of the front ions tends to be
large when the plasma temperature is low
since the electric field becomes small as the temperature
increases.

In summary, we have examined charge separation effects
in expanding electron-ion plasmas
in the context of the
diamagnetic relativistic pulse accelerator (DRPA).
When the magnetized electron-ion plasma
expands into a vacuum with no external magnetic field, only the
surface electrons are efficiently accelerated by the DRPA.
Because of the difference of the mobility between electrons
and ions, charge separation occurs and generates a strong
electrostatic field.
This electric field
accelerates the front ions and
decelerates some of the front electrons.
The initial plasma temperature affects
the charge separation effect.
As the temperature increases,
the electric field caused by the charge separation becomes
small.
Therefore, electrons in high temperature plasmas are
energized by the DRPA more efficiently 
compared to electrons in
low temperature plasmas.
On the contrary, ions are energized by charge separation
more efficiently in the low temperature case.
%
\acknowledgments

The work of KN and SPG
was performed under the auspices of the U. S. Department
of Energy (DOE) and was supported by the DOE Office of Basic
Energy Sciences, Division of Engineering and Geosciences, the
LDRD Program at Los Alamos,
and the Sun-Earth Connections Theory Program of NASA.
EL was supported by NASA Grant NAG5-7980 and LLNL Contract
B510243.



%
\end{document}